**Preprint Manuscript**

The full article is available at:

https://www.sciencedirect.com/science/article/pii/S0360835219304139

Please cite this article as:

Longo, F., Nicoletti, L., Padovano, A., d'Atri, G., Forte, M., Blockchain-enabled supply chain: an experimental study, Computers & Industrial Engineering (2019), doi: https://doi.org/10.1016/j.cie.2019.07.026

Supplementary data files are available in the journal version.

# Blockchain-enabled supply chain: an experimental study

*Francesco Longo\* (a), Letizia Nicoletti (b), Antonio Padovano (c), Gianfranco d'Atri (d), Marco Forte (e)*

(a)(c) DIMEG , University of Calabria, Arcavacata di Rende (CS) 87036, Italy
(b) Cal-Tek S.r.l., Rende (CS) 87036, Italy
(d)(e) Department of Mathematics and Computer Science, University of Calabria, Arcavacata di Rende (CS) 87036, Italy

(a)f.longo@unical.it, (b)l.nicoletti@cal-tek.eu, (c)antonio.padovano@unical.it, (d) datri@mat.unical.it, (e)marcuzzuhp@gmail.com

\* Corresponding author

**Abstract**

Despite Information and Communication Technologies (ICT) have reduced the information asymmetry and increased the degree of interorganizational collaboration, the companies participating a supply chain are less inclined to share data when information is sensible and partners cannot be fully trusted. In such a context, Blockchain is a decentralized certificate authority that may provide economic and operational benefits but companies operating in a supply chain claim to have little knowledge about Blockchain due to its novelty and to the lack of use cases and application studies.

In this work, a software connector has been designed and developed to connect an Ethereum-like blockchain with the enterprises' information systems to allow companies to share information with their partners with different levels of visibility and to check data authenticity, integrity and invariability over time through the blockchain, thus building trust. In order to explore the potential of deploying the blockchain in a supply chain, a simulation model has been developed to recreate the supply chain operations and integrated with the blockchain through the same software connector to carry out a scenario statistical analysis. Application results shows how blockchain technology is a convenient instrument to overcome collaboration and trust issues in a supply chain, to increase the supply chain overall performance, to minimize the negative consequences of information asymmetry over the echelons of a supply chain but also to discourage companies from any misconduct (e.g. counterfeiting data or low data accuracy).

**Highlights**

- Blockchain as the key for information sharing and trust issues in a supply chain.
- A real Ethereum-like blockchain network was coupled with a simulated supply chain.
- Only companies that share information achieve a significantly improved performance.
- The benefits of a blockchain-enabled supply chain exceed its costs by far.
- Blockchain drives competing supply chain companies to share data and information.

# 1. Introduction

Extant literature has widely emphasized that trust and information sharing are beneficial to supply chain performance, especially in the context of a global market with an increased tendency to outsource strategic operations (Kasemsap, 2017) to achieve greater flexibility and resilience. It has been shown that trust is also a significant predictor of supply chain's performance and fosters cost reductions, higher flexibility and better relational governance (Kim & Chai, 2017; Lee, Kim, Hong, & Lee, 2010; Singh & Teng, 2016). As can be observed in Viet, Behdani, & Bloemhof (2018), when it comes to analyzing trust and information sharing in the supply chain, studies commonly focus on demand and inventory data. Every player in a supply chain needs to forecast its customers' demand timely and accurately for its own production planning, inventory control and material requirement planning activities (Tsanos & Zografos, 2016). Any forecast uncertainty would propagate through the supply chain and amplify the order-quantity variability as we move further up the supply chain. It will eventually lead to a greater variance of production exceeding the variance of sales, excess safety stock, increased logistics costs and inefficient use of resources. This phenomenon, well-known in literature as bullwhip effect (Lee, Padmanabhan, & Whang, 1997), has been widely stemmed over the last years by the use information and communication technologies (ICT) and big data (Hofmann, 2017) that increase collaboration and data visibility. Companies have started to give partners visibility to business data (e.g. inventory data) by giving them user credentials to access their enterprise information system according to a Software-as-a-Service (SaaS) paradigm. However, information sharing and interorganizational collaboration get complicated when customers and suppliers are spread over several countries (Shore, 2001) or they do not trust each other. As supply chains become more demand-driven, data accuracy is crucial and organizations perceive trust as a vital factor of their competitive performance. However, in a low-trust scenario, supply chain partners are often averse in providing information to the other partners as they increasingly see themselves as competing entities for revenue rather than partners (Myers & Cheung, 2008), especially companies on the same echelon (e.g. wholesalers with other wholesalers, retailers with other retailers). Building trust is a slow process that requires a certain amount of accurate information to be shared for a long time by collaborating parties (Özer, Zheng, & Chen, 2011) but, on the other hand, access to accurate enterprise data and information in a supply chain is only possible when a high level of trust between the parties already exists (Ebrahim-Khanjari, Hopp, & Iravani, 2012). Despite ICT has reduced the information asymmetry and increased the degree of interorganizational collaboration (Shi, 2007), significant investments in ICT infrastructures are still required (Zhong, Newman, Huang, & Lan, 2016). Furthermore, even if companies have access to the supply chain partners' data, trust issues still exist. Indeed, companies might mislead deliberately or unconsciously the supply chain partners with inaccurate, wrong or counterfeit information that does not reflect the real data. Building trust in the supply chain leveraging on conventional ICT is therefore an expensive and long process (Poppo & Zenger, 2002), which does not always bring benefits.

Our present work proposes the use of the Blockchain as a method to provide the state of truth and trust for the information exchanged between the actors of the supply chain. A blockchain is a distributed ledger of a chronological chain of records in the form of encrypted blocks made up of all transactions executed by the participants. In the blockchain, systems can directly communicate with one another: each system can use a pair of private/public key to be identified and the communication between the systems is secure because each communication is signed by the private key of the sender (Reyna, Martín, Chen, Soler, & Díaz, 2018). Each actor of a supply chain can have a wallet in it that can be used to certify the authenticity, the integrity and invariability of data through the hash sum that is public on the blockchain and accessible at any time, while the original data are stored off-chain and exchanged between companies by using conventional methods. Therefore, blockchain creates transparency and provides a single and secure point of truth (Tapscott & Tapscott, 2016).

The Bitcoin blockchain has been the first successful application (Wattenhofer, 2016) and, today, it is one of the most widely known. However, as of June 2018, there are at least 50 different blockchains with a market value of more than 100 million USD (e.g. Ethereum, Binance, Coin, EOS, Stellar, Litecoin, Cardano, Bitcoin, TRON, Monero). Many of them are just minor modifications of the original protocol, but new ideas have also been introduced to overcome limitations on scalability and updating time of the first blockchains.

While the rise of blockchain in finance has been extremely rapid, supply chain managers, researchers and practitioners are taking longer to recognize the impact that blockchain may have on their business (Hackius & Petersen, 2017). Blockchain technology has been proved to be successful when the object under consideration has a significant value for people – e.g. food (Tian, 2016) or money – especially if the players operating in the environment where this object is created, transformed and used, do not trust each other. Although Blockchain has started to offer large benefits to make paperwork processing easier, to identify counterfeit products, to facilitate item traceability (Tian, 2016) and to operate the Internet of Things, enterprises – especially small and medium-sized companies – claim to have little knowledge about Blockchain (Kersten, Seiter, See, Hackius, & Maurer, 2017). This is mainly due to the novelty of the technology but also to the lack of use cases and application studies in literature that show Blockchain's potential benefits (Yli-Huumo, Ko, Choi, Park, & Smolander, 2016) for companies operating in a supply chain.

Supply chain management research on Blockchain is still in its infancy so it is worth to look into possible applications that may convince supply chain managers to adopt this technology as certification agent for shared data and information. A proof of concept is needed to show quantitatively which are the benefits companies could achieve in a blockchain-enabled supply chain environment before its implementation in a real context. However, data that companies send to the blockchain can be still counterfeited or be inaccurate in advance: in this sense, the trust in the blockchain still depends on the trust in partner companies. Evidence is needed to prove that actors in a supply chain that share inaccurate or counterfeit data about demand and inventory on the blockchain will not be able to achieve a high performance.

**1.1. Contribution of the study**

This research work addresses the mentioned gap in literature and industrial practice and provides in Section 2 a methodological framework to assess the performance of a blockchain-enabled supply chain. The contribution of this article is twofold. The first contribution of our research work is represented by the design and development of a software connector module that bridges an Ethereum-like blockchain with a generic enterprise information system to enable the companies to send data to the blockchain and check the data authenticity, integrity and invariability over time. In the context of this work, an Ethereum-like public blockchain, called UnicalCoin, which represents the decentralized ledger where all the information regarding demand forecasts and inventory levels are stored, has been used. In order to show quantitatively the benefits that companies can achieve in a blockchain-enabled supply chain, a supply chain simulation model has been developed to carry out "what-if" scenario analysis. It recreates the network of suppliers, carriers, wholesalers, retailers and customers, the flow of goods and information among them and all the main organizational, production and delivery processes (e.g. inventory management, demand forecasting, procurement, customers' orders arrival, deliveries etc.). The model has been integrated with the blockchain via the same software connector through REST web services to serve as a replica of a real supply chain. In this model, simulated companies can send data to the blockchain and check at their convenience the authenticity, integrity and invariability of data shared with them by other companies by using the software connector services.

The present article will eventually answer two research questions:

1. is blockchain a convenient instrument for companies operating in a supply chain - i.e. do the economic or operational benefits of building trust in the supply chain through a blockchain exceed its costs?
2. do the benefits deriving from a blockchain-enabled supply chain encourage companies not to send counterfeit or inaccurate data to the blockchain?

The second contribution of this article is indeed represented by the application study that couples the blockchain with a supply chain simulation model to quantitatively assess the benefits and advantages companies can achieve. The application study has been set up to assess the benefits of the use of blockchain for wholesalers and big-box retailers in a simulated global supply chain with low trust among the companies. Summary results are presented in Section 3 while full data from the scenario analysis are available in the supplementary file provided along with this article. Section 4 goes over the results and highlights the significant economic and operational benefits that companies can achieve by sharing accurate information with their suppliers, while no significant economic and operational benefits can be observed for those companies that just use those data. Results prove that blockchain technology is a convenient instrument to overcome collaboration and trust issues in a supply chain, to minimize the negative consequences of information asymmetry over the echelons of a supply chain but also to discourage companies from any misconduct (e.g. counterfeiting data or low data accuracy).

## 2. Materials and methods

The methodological framework here proposed to assess the performance of a blockchain-enabled supply chain consists of three components:

1. the Ethereum-like public blockchain, called UnicalCoin, which represents the decentralized ledger where all the information regarding demand forecasts and inventory levels are stored;
2. the supply chain simulation model that recreates the network of companies, the flow of goods and information among them and all their main organizational, production and delivery processes;
3. the software connector that bridges the blockchain with the supply chain simulation model (or with the enterprises' information system if it is deployed in a real supply chain environment).

### 2.1. UnicalCoin: an Ethereum-like blockchain

The choice of a specific blockchain to use is a critical step and requires the analysis of its overall reliability and the cost of using it. In this paper and for the purpose of our tests, we refer to the Ether-like blockchain UnicalCoin. UnicalCoin is an experimental Ether-like (Ethereum-like) blockchain developed at the University of Calabria (Unical), in which it is relatively easy to mine new blocks by using the Ucal cryptocurrency (UnicalCoin blockchain's cryptocurrency). This blockchain has been used only for testing purposes but, in a real environment, it would be preferable to use the official network of Ethereum. Among all the blockchains, Ethereum is one of the few that supports smart contracts, which allow to perform small and simple programs on the blockchain within the EVM

(Ethereum Virtual Machine). These programs are executed in all the machines (nodes) of the blockchain network, thus ensuring the correctness of the result that cannot be altered from one or few nodes. Participants in the UnicalCoin blockchain keep this ledger in sync through a consensus protocol (it can only be appended to, but not edited) and the higher is the number of participants, the higher is the network success. Indeed, a high number of participants is essential to ensure the immutability of the data. UnicalCoin has a limited (but relevant) number of participants (it currently counts thousands of nodes) and enabled us to replicate the Ethereum blockchain to a certain extent. UnicalCoin is not free from the risk of software attacks nor there is any guarantee to be maintained working whatever may happen. Indeed, such kind of problems exist for any permission-less chain (whoever has the right to create a personal address and begin interacting with the network), while the private ones have their own governance.

*2.1.1. Blockchain as certification agent for off-chain stored supply chain data*
A blockchain is not designed to store a big amount of data and cannot replace traditional databases. The proposed solution is to store the original data in an off-chain data storage and then publish on-chain only the hash sum of the data using the smart contract that implements the required functions. On a public blockchain, everyone can have a wallet and access the hash sum archived on the blockchain. The hash sum represents a unique key that maps to a certain data or document and the use of a cryptographic hash function allows one to easily verify the origin and authenticity of data by checking whether some data map onto a given hash value in the blockchain. However, if the original data is unknown, the hash sum does not map back to it. On one hand, companies should grant a partner of theirs access to data stored off-chain; on the other hand, the partner can verify the accuracy and validity of data as the hash sum calculated from the original data should correspond to the hash sum archived in the blockchain. Therefore, the choice to store the original data off-chain and use the blockchain as certification agent guarantees the privacy and regulated access to confidential data as well as data integrity. In this study, the hash function chosen is SHA-512 (Gueron, Johnson, & Walker, 2010). A common problem of the hash function is the collision, namely the possibility that two different inputs have the same hash sum. The possibility of a collision won't be an issue given the fact that the system does not use the hash to recognize the document but only to verify that the information stored off-chain and shared between partners of the supply chain is the same as the one hashed initially.

As off-chain technology, it is possible to choose between different solutions, such as distributed and decentralized technologies, very popular in blockchain contexts (such as IPFS and SWARM), or more conventional databases. IPFS (InterPlanetary File System) is an innovative peer-to-peer technology for distributing files on the web (Benet, 2014), which should increase file transmission performance and increase security as the files are distributed on different nodes. The file transmission is based on the BitSwap protocol, which is based in turn on the well-known BitTorrent protocol. Distributed hash table (DHT) is used in IPFS for routing and file addressing, because the files are indexed by the hash sum of the content. File versioning is instead managed with the same techniques as the GIT protocol. SWARM is also a distributed storage platform, which was born as a support tool for the data distribution on the Ethereum blockchain (Trón, Fischer, Nagy, Felföldi, & Johnson, 2016). Since SWARM was born within the Ethereum ecosystem, its integration with the blockchain is very deep (for example, a payment service is provided for storage via Ether and several other integrations via smart contracts). Today, SWARM has become a distributed and decentralized storage technology that can work independently of Ethereum blockchain, so it is able to store files that are not present on the Ethereum blockchain.

In general, IPFS and SWARM are very similar technologies as both are peer-to-peer distributed storage systems where the files are indexed through the hash of their content. They adopt a decentralized transfer system, offer low-latency performance, they are fault-tolerant and censorship-resistant. The differences mainly concern the used protocol and the peer management techniques.

If we use these technologies, the files can be consulted by anyone in the network. This implies that, before being distributed, due to the confidential nature of data, the files should be encrypted. This is actually not necessary as, once the files are created or available on the enterprise information systems, their immutability and accuracy is guaranteed by the blockchain. The files are indexed by the hash of their content, so if the file will be modified, its hash will change too.

Furthermore, despite both systems implement different mechanisms to preserve low-demand files, none of the two technologies guarantees that the file will always be available in the system. If the companies of a consortium or of a supply chain want to adopt one of these technologies, they could mitigate the file persistence problem by setting up one or more nodes in the peer network, which are controlled by the companies and could preserve the files of their interest indefinitely.

However, in order to overcome the above-mentioned issues, classic relational databases are a preferred way over innovative storage techniques to store supply chain data and documents off-chain. Enterprises' information systems usually rely on these databases that also offer good performance on small-size data like in the case of this application. For the purpose of this analysis, one of the most common relational database management systems, MySQL, was used to store data off-chain.

*2.1.2. Blockchain costs*

A transaction fee has to be paid whenever a blockchain participant tries to execute a transaction. Each transaction in the public ledger is verified by a majority of participants in the system through a consensus mechanism. In order to have the transaction processed, certain computers in the network, referred as "miners", should find an eligible hash for their block of transactions and this process is a computationally intensive problem. As mentioned earlier, the higher is the number of participants in the blockchain, the higher is network reliability and success, therefore miners must be rewarded for their work (appending a new block of transactions to the blockchain). A transaction fee is assigned to the miner that created the block where the transaction has been added. The fee is calculated in the blockchain cryptocurrency and each blockchain has its own method for calculating this cost. The Bitcoin blockchain calculates the cost of a transaction based on the space it occupies within the block. The cost for the bitcoin blockchain is 11 satoshi (the smallest bitcoin unit of measurement) per byte stored. The Ethereum blockchain calculates the cost based on the gas used by a transaction. In general, we could define the gas as the cost of each operation performed on the blockchain, from a simple ether transaction to the execution of a smart contract's function. The amount of gas is directly proportional to the difficulty of the operation to be performed. Also, the storage of data on the blockchain has a cost in terms of gas. The cost of gas for each operation is defined and hard-coded in the Ethereum blockchain's software (Wood, 2019). When a transaction is submitted, a given amount of gas is associated to it. An Ether value is assigned to a unit of gas, so the total transaction cost C is defined as:

$$C = G \times P \quad (1)$$

where $G$ is amount of Gas and $P$ is the Gas Price in Ether.

The amount of gas necessary to perform the transaction that stores the hash on the blockchain is constant, while the gas price (in terms of ether) depends on the blockchain participants. The gas price is a key factor as it determines the time required by a transaction to be mined. The reward of a miner in the blockchain is finally given by a constant value related to the block and the sum of the costs of the transactions included into the block (which depends on the amount of gas and on the gas price). Therefore, the miners would choose the transactions with the highest reward to be added to their block. For this reason, if this system will be used in a real supply chain context, an evaluation of the gas price for a single transaction in the Ethereum blockchain is necessary. The amount of gas is constant and is 190.000 while the gas price change proportionally with the Ethereum network traffic (in a situation of congestion, the gas price is very high). To make a good evaluation of the transactions cost, we retrieved the average gas price from https://etherscan.io/chart/gasprice, the average ether price from https://etherscan.io/chart/etherprice and we evaluated the total transaction cost considering the daily data. For our application, we assume the average of the cost of a transaction to be on average 1,09$ (0,93€), with a maximum of 22.7$ (19,40€) and a minimum of 0.01$ (0,01€). The transaction cost mainly depends from the smart contract structure. The structure used for the present application can be simplified to be cheaper than the current one, therefore the actual cost of using the blockchain can be even lower than the one considered as an estimate in this study.

Every time a blockchain participant (one of the members of the supply chain) submit a transaction to the blockchain network, the transaction will be pending until a network node put it into a block to be mined.

## 2.2. A supply chain simulation model

In order to assess quantitatively the benefits that companies can achieve in a blockchain-enabled supply chain, a supply chain simulation model has been developed to carry out "what-if" scenario analysis. It recreates the network of suppliers, carriers, wholesalers, retailers and customers, the flow of goods and information among them and all the main organizational, production and delivery processes (e.g. inventory management, demand forecasting, procurement, customers' orders arrival, deliveries etc.). In this section, the supply chain operations are first conceptualized and then implemented in a multi-paradigm (discrete-event and agent-based) Java-based simulation model developed from stcratch. Some mathematical models underpinning the data forecasting and inventory management are presented as they are deemed to be important to understand the type of data and information shared between the actors of the supply chain and how the data accuracy and authenticity may impact on the enterprises' performance.

### 2.2.1. The supply chain conceptual model

In our global supply chain conceptual model, a single network node can be considered as a wholesaler or a big-box retailer, which operates a series of physical stores. In the case of this study, three wholesalers and twenty big-box retailers have been modelled as depicted in Figure 1 (four manufacturers are also considered in the present supply chain conceptual model but their description is out of the scope of this paper). As the stores are operated by the same entity (the big-box retailer), we assume real-time data transparency and visibility among them and no trust issues within the same organization. The big-box retailers are in competition among themselves as they sell 60 homogeneous products in the same regional markets. Starting from the end of the supply chain, customers' market demand at the stores of the big-box retailers can be modelled with a Poisson process. The arrival process is supposed to be independent for every item and the quantity required for each item can be modelled as a triangular probability distribution with different levels of intensity and variability. Once the customer arrives at the store, the quantity is compared with the inventory and, if possible, the order is satisfied (otherwise lost demand is recorded for fill rate calculation). The inventory level is checked before the business hours and, in case a purchase order is required, the big-box retailer can choose its supplier, i.e. the wholesaler. Such decision is made considering the lead time, the lead time demand and the quantity immediately available at the wholesalers. In this study, the lead time demand is evaluated by using a single exponential smoothing (SES), while the quantity that the big-box retailers eventually receive can be slightly different from the quantity ordered by the retailers due to problems at the manufacturers' sites or at the wholesalers' sites. Every day the wholesalers try to satisfy purchase orders with the same priority. If the inventory level of an item is not enough to satisfy the retailers demand, the available quantity is divided among the retailers considering the ordered quantity as weighting factor. Lost quantities are recorded so that the wholesaler's fill rate can be evaluated. Once per day, the wholesaler checks the inventory level for each item and request a certain amount of products to the manufacturers. The wholesaler's order waits in a queue until it is processed and the products delivered to the wholesaler. For the sake of completeness, each plant is modelled as a group of machines and each machine can manufacture all the type of items (with different efficiency rates, working times and setup times when switching from a product to another), however their description is not part of this work.

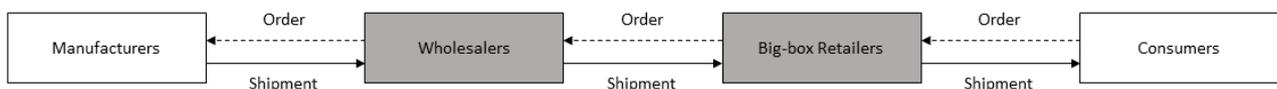

Figure 1. A two-stage supply chain conceptual model

2.2.2. Simulation model and data

After the supply chain has been conceptualized, a multi-paradigm (discrete-event and agent-based) supply chain simulation model has been developed to recreate the supply chain above described. A specific Java class has been developed for every agent in the supply chain (products, manufacturers, wholesalers, big-box retailers with their stores and customers). A discrete-event based approach has been adopted to implement organizational processes, such as inventory control, order fulfillment, etc. A Java class has been also created to represent purchase orders. All the simulation events (e.g. beginning of inventory control checking process) are generated using event generator objects and, in correspondence of such events, specific methods elaborate and update the information about demand forecasts and inventory levels stored in database tables. Following this modeling approach, we have obtained a flexible, parametric and time efficient simulation model.

Orders arrive at each store of the big-box retailer $j$, $j = 1,\ldots,20$ with an interarrival time distributed according to a negative exponential function. Stores use a Re-Order Level - Target Level (r, R) policy to manage their inventory and they also use a Single Exponential Smoothing (SES) to predict the demand. The probability distribution for the quantity ordered by the customers at every store of the big-box retailer is assumed to be triangular. Data and parameters for setting up the big-box retailer's stores in the simulation model are given in Table 1.

*Table 1. Big-box retailer's parameters*

| | |
|---|---:|
| Average order inter-arrival time [sec] | 5000 |
| Lower bound for the inter-arrival time [sec] | 3600 |
| Upper bound for the inter-arrival time [sec] | 7200 |
| Lead time (days) | 5 |
| Parameter 1 for SES (interval for historical data) [days] | 10 |
| Parameter 2 for SES (alfa) | 0,7 |
| Standard deviation factor (safety time) | 2 |
| Standard deviation of the lead time (safety time) | 0,5 |
| N (number of day for SS) | 20 |
| Review Period S (for r, R policy) | 3 |
| Triangular Distribution – Minimum Value | 18 |
| Triangular Distribution - Mode | 30 |
| Triangular Distribution – Maximum Value | 44 |
| Ordering Cost [€/order] | 15 |
| Transportation Cost [€/order] | 60 |
| Reception Cost [€/order] | 25 |
| Storing Cost [€/item] | 0,65 |
| Obsolescence Cost [€/item] | 0,15 |
| Deterioration Cost [€/item] | 0,15 |
| Interest Cost [€/item] | 0,05 |

Every wholesaler $i$, $i = 1,2,3$, is characterized by a Dynamic Safety Stock as inventory management policy and a Single Exponential Smoothing (SES) as forecasting method. The parameters of such methods as well as the costs per single order and per single item are summarized in Table 2.

*Table 2. Wholesaler's parameters*

| | |
|---|---:|
| Lead Time [days] | 1 |
| Parameter 1 for SES (interval for historical data) [days] | 15 |
| Parameter 2 for SES (alfa) | 0,6 |
| Standard deviation factor (safety time) | 2 |
| Standard deviation of the lead time (safety time) | 0,5 |
| N (number of days for SS) | 20 |
| Review Period S (for r, R Policy) | 2 |
| Ordering Cost [€/order] | 20 |
| Transportation Cost [€/order] | 100 |
| Reception Cost [€/order] | 40 |
| Storing Cost [€/item] | 0,4 |
| Obsolescence Cost [€/item] | 0,15 |
| Deterioration Cost [€/item] | 0,15 |
| Interest Cost [€/item] | 0,05 |

For the purpose of this study, we describe in the following the inventory control and demand forecasting policies used in the simulation model as they are crucial to show the type of information that is shared between big-box retailers and wholesalers and to understand how it impacts on the enterprise's performance. Further details about the implemented inventory control policy can be found in Longo & Mirabelli (2008).

*2.2.3. Inventory control and demand forecasting models*

The wholesalers and the stores use a modified continuous review policy (r, R) as inventory control policy to calculate the time when a purchase order has to be emitted and a certain quantity to be ordered. The quantity of item $k$ to be ordered at time $t$ at the network node $i$, $q_{ik}(t)$, depends on the target level $\theta_{ik}(t)$ and the inventory position, $\pi_{ik}(t)$:

$$q_{ik}(t) = \theta_{ik}(t) - \pi_{ik}(t) \tag{2}$$

The target level $\theta_{ik}(t)$ is given as the sum of the safety stock at time $t$ of the item $k$ at the network node $i$, $ss_{ik}$, and the lead time demand of the item $k$ at network node $i$, $\delta_{ik}(t)$ - here evaluated by using the single exponential smoothing as described in (2):

$$\theta_{ik}(t) = \delta_{ik}(t) + ss_{ik}(t) = \sum_{r=t+1}^{t+lt_{ik}} \varphi_{ik}(r) + ss_{ik}(t) \tag{3}$$

where $\varphi_{ik}(r)$ is the demand forecast at time $r$ of the item $k$ at the network node $i$ and $lt_{ik}$ is the lead time of the item $k$ at the network node $i$.

The inventory position, $\pi_{ik}(t)$, is instead given by the on-hand inventory, $\alpha_{ik}(t)$, the quantity already on order, $\beta_{ik}(t)$, and the quantity to be shipped, $\gamma_{ik}(t)$:

$$\pi_{ik}(t) = \alpha_{ik}(t) + \beta_{ik}(t) + \gamma_{ik}(t) \tag{4}$$

The time for a purchase order to be emitted can be calculated therefore starting from this condition:

$$\pi_{ik}(t) < \lambda_{ik}(t) = \theta_{ik}(t) = \delta_{ik}(t) + ss_{ik} \tag{5}$$

where $\lambda_{ik}(t)$ is the re-order level at time $t$ of the item $k$ at the network node $i$ and the safety stock $ss_{ik}(t)$ is calculated as standard deviation of the lead time demand $\delta_{ik}(t)$. The re-order level, the target level and the safety stock are supposed to be constant over the review period $\rho$. If we indicate the demand forecast over $\rho$ as $\tau_{ik}(t)$ at a specified time $t$. Therefore, we can write:

$$\lambda_{ik}(t) = lt_{ik} \times \frac{\tau_{ik}(t)}{\rho} + ss_{ik} \qquad (6)$$

$$\theta_{ik}(t) = lt_{ik} \times \frac{\tau_{ik}(t)}{\rho} + \lambda_{ik}(t) \qquad (7)$$

In addition, we define the total cost for a purchase order emission (POE) and the total cost for storage (ST) as respectively:

$$TC_{POE,ik} = C_{ik,o} + C_{ik,t} + C_{ik,r} \qquad (8)$$

$$TC_{ST,ik} = C_{ik,st} + C_{ik,w} + C_{ik,ob} + C_{ik,i} \qquad (9)$$

where:
- $C_{ik,o}$, order placing cost for item $k$ at the network node $i$;
- $C_{ik,t}$, transportation cost for item $k$ at the network node $i$;
- $C_{ik,r}$, order reception cost for item $k$ at the network node $i$;
- $C_{ik,st}$, storage cost for item $k$ at the network node $i$;
- $C_{ik,w}$, worsening cost for item $k$ at the network node $i$;
- $C_{ik,ob}$, obsolescence cost for item $k$ at the network node $i$;
- $C_{ik,i}$, interest cost for item $k$ at the network node $i$.

The optimized review period, $\bar{p}_{ik}(t)$, can be calculated as the argument that minimizes, on the basis of the demand forecast, the unitary inventory cost $ic_{ik}(t)$:

$$\bar{p}_{ik}(t) = arg\, min(ic_{ik}(t)) = arg\, min\left(\frac{TC_{POE,ik} + TC_{ST,ik} \times \sum_{t}^{t+T-1}(t-1) \times \varphi_{ik}(t)}{\sum_{t}^{t+T-1} \varphi_{ik}(t)}\right) \qquad (10)$$

If we indicate with $\bar{\tau}_{ik}(t)$ the forecast of the demand over the optimized review period $\bar{p}_{ik}(t)$, the target level can be reformulated as:

$$\theta_{ik}(t) = \bar{\tau}_{ik}(t) + \lambda_{ik}(t) \qquad (11)$$

In other words, $\bar{\tau}_{ik}(t)$ is the optimal lot size calculated by means of the demand forecast.

*2.2.4. Supply chain performance measures*

In terms of supply chain performance measures, the simulation model calculates the orders' fill rate, the on-hand inventory, the total inventory costs and the average inventory cost per day and per single item, revenues, costs and net profit. The fill rate is calculated both for the wholesalers and the big-box retailers, just after the end of the business hours, as the ratio between the number of fully satisfied orders, $FSO_{ik}(t)$, and the total number of orders, $TO_{ik}(t)$, as expressed in (12).

$$FR_{ik}(t) = \frac{FSO_{ik}(t)}{TO_{ik}(t)}, \qquad \forall i, k, t \tag{12}$$

The on-hand inventory is monitored before and after the business hours providing, for each day, the average on-hand inventory. The total inventory cost can be easily calculated considering (8) and (9) and the purchase cost (the price $p_{ik}$ times the quantity received $q_{ik}(t)$) as reported in (13):

$$TIC_{ik}(t) = TC_{POE,ik} + TC_{ST,ik} * \alpha_{ik}(t) + p_{ik} * q_{ik}(t) \tag{13}$$

**2.3. A software connector to bridge the simulation model with blockchain**

In order to connect the UnicalCoin blockchain with the enterprises' information systems and, for the purpose of application study, with the simulation model, a software connector has been developed. The application has been implemented in Java because:

- it is very mature and widely used in open-source contexts;
- there are many Java libraries for Ethereum, not only for the interaction with a node, but also for the generation of smart contract wrappers,
- most enterprises' information systems are Java-based, which makes the developed connector suitable to several real companies;
- the supply chain simulation model is also Java-based, which means that the same connector could be used to bridge also the simulation model with the blockchain for the purpose of our application study.

The application's general framework is depicted in Figure 2 and applies to the connection between the blockchain and the supply chain simulation model (the same approach can be used with the real companies though). It includes the following objects:

- the Company, which has an Ethereum wallet (i.e. an address) that guarantees the data authenticity and represents the (simulated or real) supply chain member allowed to publish its own data or read the data of other companies;
- the SharedInfo, which is the instance of data shared by the company;
- the TransactionVerification, which contains all the data related to an Ethereum transaction (e.g. the block number or transaction hash) when a SharedInfo is sent to the application to be shared and stored into the blockchain.

Supply chain data are stored off-chain in MySQL databases, whereas interactions with the blockchain to store the hash sum of this data are regulated by the services provided by the software connector. Services cannot be executed in every blockchain but they require the support of smart contracts (Xu et al., 2016) and have been grouped into three categories:

- the Communication services, which represent the act of transferring data between different modules and can be easily executed with a simple transaction;
- the Coordination services, which have the task of transferring control between different modules;
- the Facilitation services, which support and optimize the interactions and are naturally present in the blockchain's technology like the transaction validation, the transaction signature, the data invariability property (implemented by design in the blockchain).

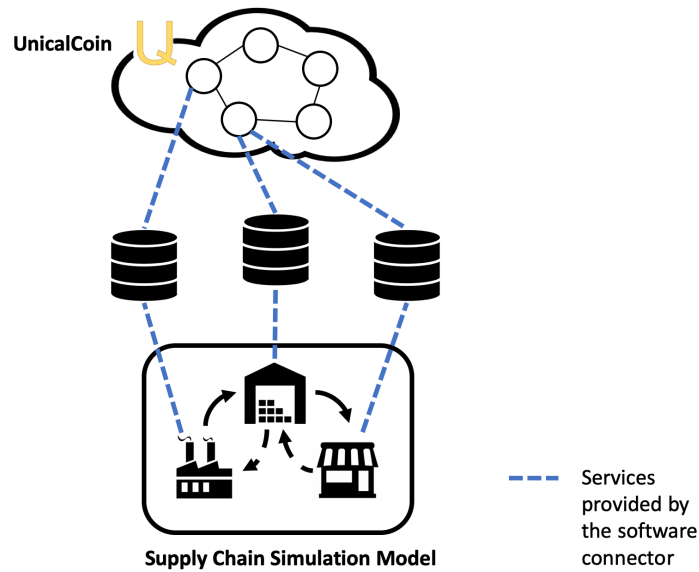

*Figure 2. The system's general framework*

Based on these services, the software connector can provide the companies with the following functionalities (see Figure 3):

i. to allow or deny other company's Ethereum address to publish and access data;

ii. to publish data on the blockchain;
iii. to search and monitor data, but only those they have been granted access to;
iv. to verify the data authenticity, integrity and invariability through the hash sum.

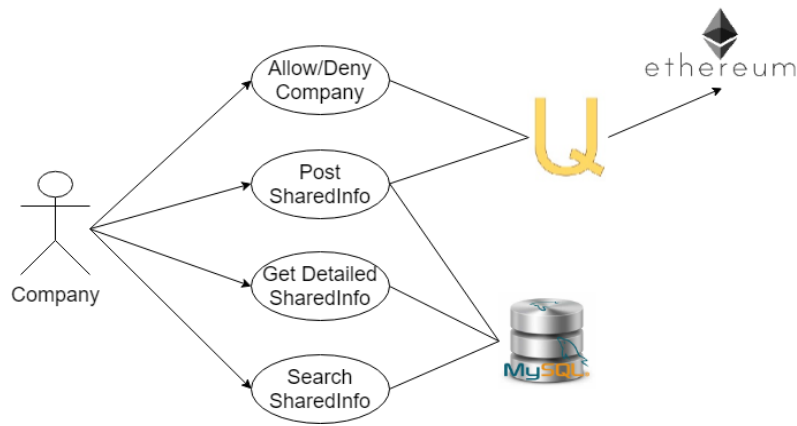

*Figure 3. Application architecture: use case diagram*

All accesses to the system are made via restless endpoints, which are protected by HTTP-Basic authentication (IETF, 2015). Authentication as a requirement for accessing data ensures that only those who are authorized can view them. More generally, access to endpoints with largely verified solutions, such as HTTP-Basic authentication with the HTTPS protocol, guarantees security and respect for privacy. Furthermore, the smart contract that implements all these functionalities is programmed to allow data to be published on the blockchain only by certain addresses, which must be first authorized. Indeed, the confidential nature of data requires data access regulation. The enabled addresses can vary over time so a mechanism is required to activate or deactivate them. An Ethereum address applies for being authorized and its request remains pending until the majority (50%+1) of the authorized addresses speak or not in favor of the authorization of the pending address by means of the "Allow/Deny Company" function.

The "post SharedInfo" function enables authorized companies to share data on the blockchain through the application. Each company can run the software connector on their own machines or, alternatively, companies can entrust the service to a third party that will store, share and publish on the blockchain on the company's behalf. The second option may seem a contradiction to the blockchain concept, but it well describes the fact that the stakeholders do not care about who actually published the data. What matters is that the hash sum of the original data is published on the blockchain and that stakeholders are able to crosscheck it with the original data. Data are published on the blockchain according to the following process depicted in Figure 4:

1. The company A (i.e. an agent in the simulation model or a real company) sends to the application a JSON file with the data to share, the reference date, and the visibility group (i.e. the companies that are authorized to view their data);

2. The application receives the data and start to process it. The SharedInfoPostServices catch the request, convert some parameters in a more useful type and throw the request to the SharedInfoServices. The SharedInfoServices performs some integrity checks on the data received, creates the hash sum of the payload, fill the bean that will be stored and throw the SharedInfoManager. It then sends the request to an UnicalCoin node by a signed transaction.
3. The UnicalCoin node receives the transaction, checks the validity and store it in the blockchain; when stored, the node responds to the application, sends the transaction verification details and the shared data id that is generated by the smart contract. Full data also include where the hash sum has been stored on the blockchain (for example, block number, transaction number etc.).
4. The application receives from the node the generated transaction's id, the verification details and eventually the application stores this data into the database off-chain.

Data can be published in a structured way with a well-defined frequency, for example once a week, once a day or every X hours. The frequency generally depends on the needs of the supply chain members: for example, in the case of the application study here proposed, lead time demand data are published every day at the end of the business hours on the blockchain.

In order to implement the "get SharedInfo" and "search SharedInfo" functions, as soon as a smart contract is executed in the blockchain and generates verification data (TransactionVerification), an event can be emitted. Events are an essential tool for development of decentralized apps and they are very useful to monitor when other companies share new data. As showed in Figure 5, an UnicalCoin node (similarly to Ethereum) allows its clients to listen to well-defined events so that every company's application registered on the UnicalCoin node will be notified in the event new data will be published. Companies can finally query other companies' databases to read data they have access to and visibility on (they are part of the visibility group). Since blockchain is public, those data may be shared with certain stakeholders who have agreed on their reading rights, thus allowing the indisputable verification of information.

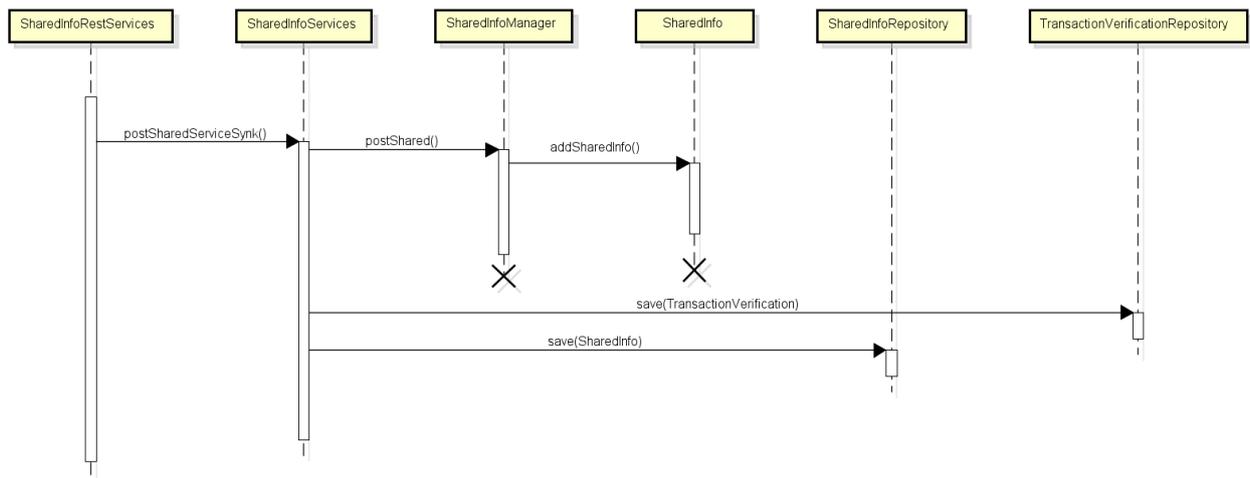

*Figure 4. Application architecture: sequence diagram*

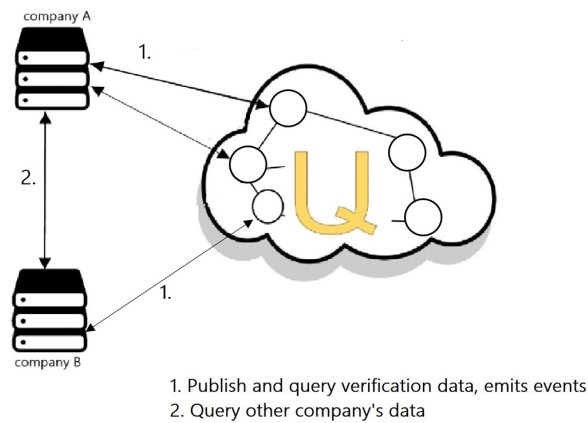

*Figure 5. Publish and query mechanisms*

In order to implement all these functionalities, REST architecture has been adopted to implement the three categories of services mentioned above. RESTful web services let the simulation model (or the enterprise information system) interoperate with the UnicalCoin blockchain. Spring Boot is the Java framework used for the development of this module, which facilitates and speeds up the development of REST solutions and allows the integration of multiple other frameworks such as Hibernate for the data persistence management.

The software was developed as a classic MVC (Model View Controller). The Model is implemented in the model package and contains all the beans, which do not contain logic but only data. The dao package contains the class that allow the interaction with the database. In particular, the model and dao packages implement Hibernate interfaces. The core package implements the business logic functions, so it can be considered as the controller. The functionalities are grouped by Java classes, one core for the SharedInfo and the other for the company. This class implements the "Services" annotations of Spring and the injection, also provided by the framework. The smart-contract package contains the wrapped smart-contract and a manager that facilitate the usage of the smart-contract. The REST package contains two classes that implements a Spring Rest Controller: each class provides all the methods needed to access to the functionalities related to the beans. This layer implements the view by exposing the WS to the internet: once a request is received, the REST layer sends the request to the logic layer that is able to process it.

## 3. Results

The UnicalCoin blockchain has been used in the context of this paper as a secure, trusted and decentralized ledger where wholesalers and big-box retailers store the hash sum of supply chain related relevant data archived off-chain. For the purpose of this study, we enabled only the big-box retailers to send a transaction to the blockchain network at the end of the business day, thus allowing the wholesalers (with specific levels of visibility) to receive quasi real-time the exact information about the market demand for the 60 items. For the purpose of this study, two scenarios are considered:

- in the first scenario, referred to as No-IS (No Information Sharing), no information about the lead time demand is shared between the big-box retailers and the wholesalers because of low-trust among the parties or, even if information is shared, companies do not trust each other and data is assumed to be useless for forecasting and planning purposes;
- in the second scenario, the information about the lead time demand is shared at the end of the business day by every big-box retailer and made available to a specific group of wholesalers. In this scenario, referred to as B-IS (Blockchain-enabled Information Sharing), wholesalers can check the data authenticity and invariability over time by using the hash sum of the data stored in the blockchain, thus providing a secure point of trust.

The simulation time in both scenarios is 60 days and has been replicated three times for each scenario in order to mitigate the effects of randomization and stochasticity on the results. Summary supply chain performance measures in terms of average over the three replications are provided in this section and statistically analyzed. For the reader's convenience, full data from the scenario analysis carried out with the simulation model are available in the supplementary file provided along with this article.

### 3.1. Blockchain performance and costs

At the moment of the experimentation with the UnicalCoin blockchain, other four applications were using this network, so the pending time is very low. To evaluate the transaction pending time using the UnicalCoin blockchain, we consider the difference between the time when the block containing the transaction is mined and the time when the request is received. The average time is estimated to be 16.3 seconds, the maximum is 146 seconds and the minimum is 2 seconds. Since the simulation length is set up to 60 days and every member of the supply chain submit a transaction to the blockchain network containing the lead time demand for each item, 60 transactions per day will be generated by each retailer of the blockchain-enabled supply chain. Therefore, we can calculate the total minimum, average and maximum blockchain transaction costs as summarized in Table 3.

*Table 3. Blockchain usage summary parameters*

|  | Min | Avg | Max |
|---|---|---|---|
| Total blockchain transaction time per node per day (sec) | 120 | 978 | 8.760 |
| Total blockchain costs for each retailer (€) | 0,765 | 83,7 | 1.746 |
| Total blockchain costs for each wholesaler (€) | 5,1 | 558 | 11.640 |
| Total blockchain transactions costs (€) | 30,6 | 3.348 | 69.840 |

### 3.2. Wholesalers' performance

In first place, average revenues (R), missing revenues (MR) due to unsatisfied demand, total costs (TC) and profit margin (PM) are reported in Table 4 for the three wholesalers calculated over the three replications. Considering that all the other costs are not influenced by the information sharing (e.g. personal costs, depreciation, utility expenses, etc.), the costs reported in Table 4 include only the Total Inventory Cost (TIC) and the costs for the use of the blockchain as they are the only costs that are subject to a potential variation due to the use of this technology (the maximum costs are considered, $BC_{max}$).

*Table 4. Wholesalers' average economic performance indicators in the two scenarios*

| Scenario | R (€) | MR (€) | TIC (€) | $BC_{max}$ (€) | TC (€) | PM (€) |
|---|---|---|---|---|---|---|
| B-IS | 87.910.850,67 | 0,00 | 37.995.012,45 | 11.640,00 | 38.006.652,45 | 49.904.198,21 |
| No-IS | 74.854.573,33 | 17.813.402,67 | 24.981.013,56 | 0,00 | 24.981.013,56 | 49.873.559,77 |

Average revenues and profit margins in both scenarios for the three wholesalers calculated over the three replications are illustrated in Figure 6 and Figure 7.

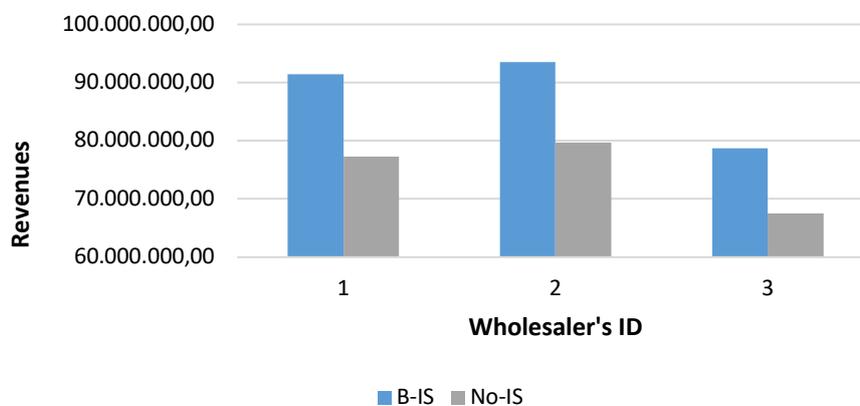

*Figure 6. Wholesalers' revenues in the two scenarios*

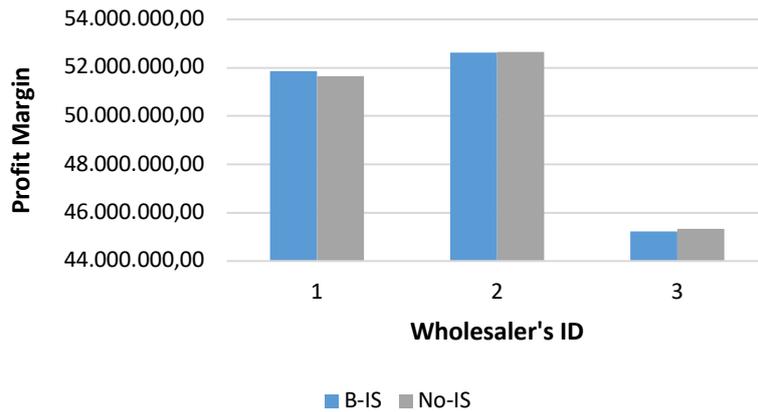

*Figure 7. Wholesalers' profit margin in the two scenarios*

Average revenues for the wholesalers increase of a 17,44% (18,32% for the wholesaler 1, 17,29% for the wholesaler 2 and 16,62% for the wholesaler 3) when we consider the B-IS scenario as opposed to the No-IS scenario. Missing revenues data also confirm that in the B-IS scenario all the orders are fulfilled by the wholesalers (then no missing revenues are observed in the B-IS scenario).

However, due to the higher costs, the average profit margin before and after the use of the blockchain does not undergo a substantial variation (49.904.198,21€ in the B-IS scenario versus 49.873.559,77€ in the No-IS scenario). Their profits do not go through any significant improvement as confirmed by the non-parametric Mann-Whitney test. In this case, $H_0$ is "the medians of the wholesalers' profits are equal" and as alternative hypothesis that "the median at B-IS is greater than the median at No-IS". The results reported in Table 5 shows that there is not enough power to reject the null hypothesis and consider that the wholesalers' profits are greater in the B-IS scenario than the No-IS scenario.

*Table 5. Non-parametric Mann-Whitney test for the wholesalers' profits*

| Scenario | N | Median | Difference | Lower Bound for Difference | W-Value | P-Value |
|---|---|---|---|---|---|---|
| B-IS | 3 | 51.852.061 | -23097,1 | -7.408.578 | 10,00 | 0,669 |
| No-IS | 3 | 51.651.761 | | | | |

A closer look at the inventory costs is therefore necessary. The average fill rate (FR) for the wholesalers in the No-IS scenario is 63,67%, while in the B-IS scenario it is 100%, meaning that all the orders have been fulfilled (see Figure 8 for a graphical comparison). Based on the inventory position (IP) of the wholesalers averaged per every item, the average inventory costs (AIC) per day and per item and the total inventory costs (TIC) are provided in Table 6 for the two scenarios.

*Table 6. Wholesalers' inventory management key outcomes*

| Scenario | FR (%) | IP | AIC (€) | TIC (€) |
|---|---|---|---|---|
| B-IS | 1,00 | 12.943,89 | 34.516,54 | 113.985.037,36 |
| No-IS | 0,64 | 9.837,82 | 30.289,94 | 100.037.281,51 |

The fill rate increases from 63,67% to 100,00%, meaning that the wholesalers never go out of stock and are able to fulfill all the orders that they receive by the retailers when the latter share trustworthy information through the blockchain about the lead time demand. A significant difference between the two scenarios can be observed from a face validation of the bar graphs representing the inventory position and the total inventory cost (see Figure 9 and Figure 10).

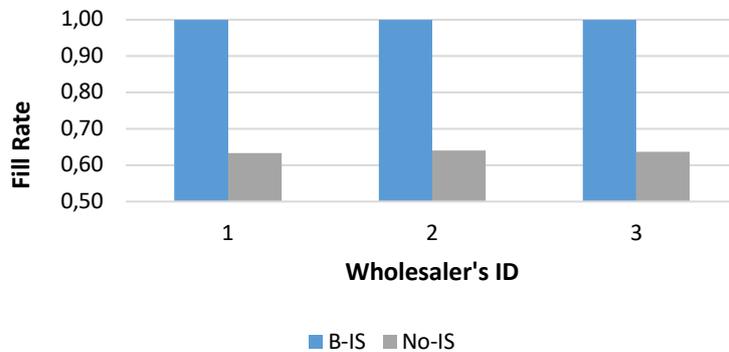

*Figure 8. Wholesalers' average fill rate in the two scenarios*

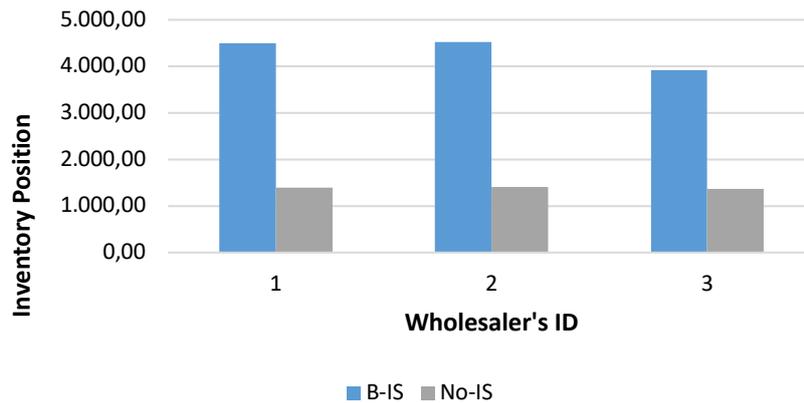

*Figure 9. Wholesalers' inventory position in the two scenarios*

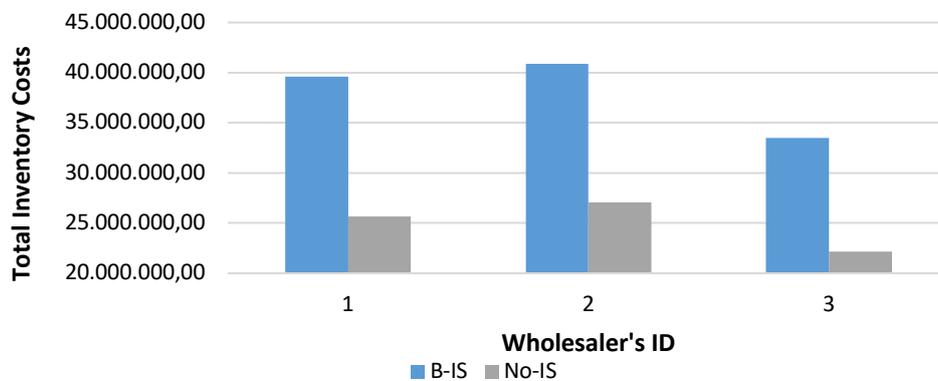

*Figure 10. Wholesalers' total inventory costs in the two scenarios*

## 3.3. Big-box retailers' performance

The same key economic performance indicators (R, MR, TIC, $BC_{max}$, TC and PM) are used to monitor the economic performance of the big-box retailers. Detailed values per each big-box retailer in the supply chain in both scenarios are illustrated in Figures 11-15, while a summary is provided in Table 7.

Table 7. Big-box retailers' economic performance indicators in the two scenarios

| Scenario | R (€) | MR (€) | TIC (€) | $BC_{max}$ (€) | TC (€) | PM (€) |
|---|---|---|---|---|---|---|
| B-IS | 17.345.682,76 | 603.835,56 | 6.242.163,80 | 1.746,00 | 6.243.909,80 | 11.101.772,96 |
| No-IS | 15.400.965,14 | 2.548.553,18 | 5.358.926,91 | 0,00 | 5.358.926,91 | 10.042.038,23 |

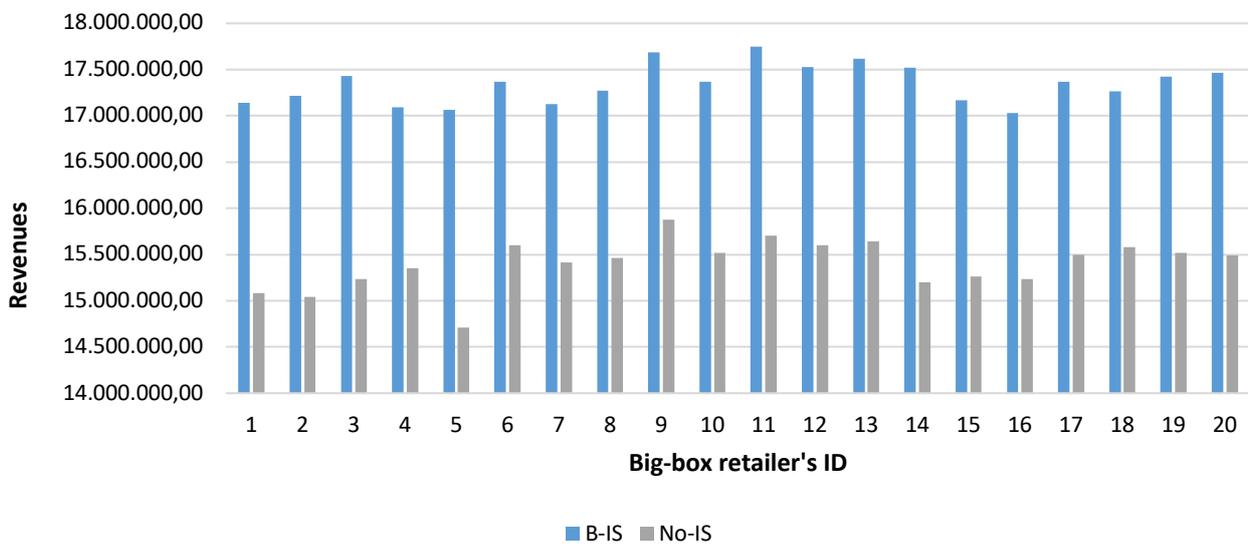

Figure 11. Big-box retailers' revenues in the two scenarios

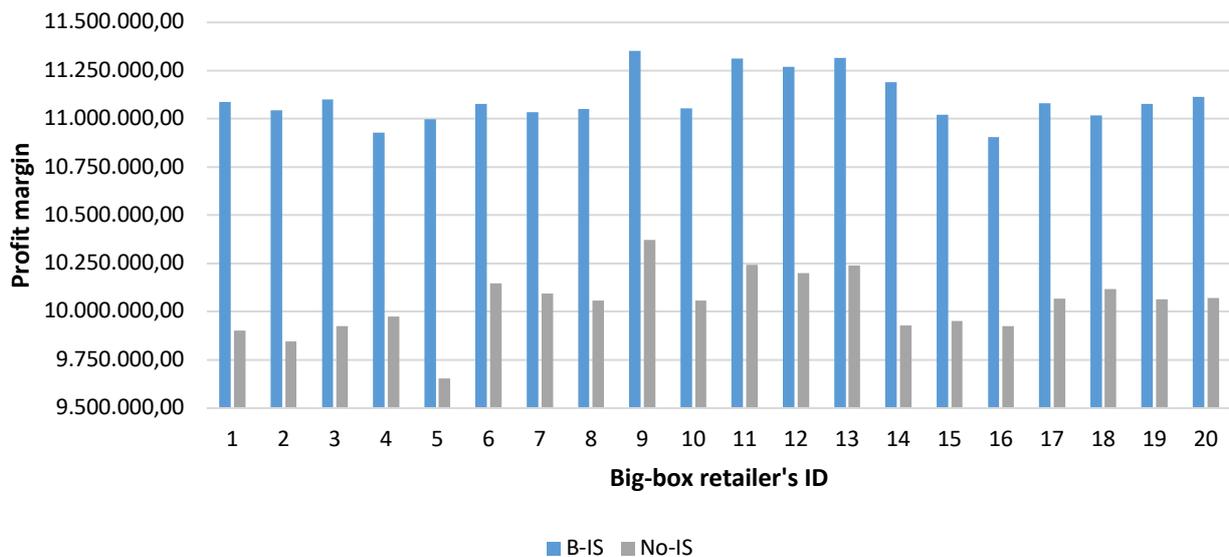

Figure 12. Big-box retailers' profit margin in the two scenarios

In the B-IS scenario, big-box retailers are able to obtain higher revenues, to minimize the missing revenues (which decreases on average by a significant 76,31%) thanks to a blockchain-enabled information sharing that guarantees trust among the parties. As final results, the profits of the big-box retailers will be significantly greater in the B-IS scenario than those in the No-IS. The difference for the retailers' profits in the two scenarios has been tested again by applying the non-parametric Mann-Whitney test, which shows that the null hypothesis "the medians of the retailers' profits are equal" should be rejected and the alternative hypothesis that "the median of the retailers' profits at B-IS is greater than the median at No-IS" is valid (results are given in Table 8).

*Table 8. Non-parametric Mann-Whitney test for the retailers' profits*

| Scenario | N | Median | Difference | Lower Bound for Difference | W-Value | P-Value |
|---|---|---|---|---|---|---|
| B-IS | 20 | 11.077.151 | 1.043.020 | 977.748 | 610,00 | 0,000 |
| No-IS | 20 | 10.062.462 | | | | |

Key indicators of the inventory management are worth of investigation to understand where these higher revenues come from. A summary table about the key inventory management indicators is provided in Table 9.

*Table 9. Big-box retailers' inventory fill rate in the two scenarios*

| Scenario | FR (%) | IP | AIC (€) | TIC (€) |
|---|---|---|---|---|
| B-IS | 0,96 | 1.106,23 | 1.733,94 | 6.242.163,80 |
| No-IS | 0,84 | 1.102,86 | 1.488,59 | 5.358.926,91 |

If we consider the three replications and all the 20 retailers, the average fill rate in the No-IS scenario is 84,13% versus a 95,83% in the B-IS scenario, with an increase of 11,70% (Figure 13 shows the average fill rate over the three replications for the retailers).

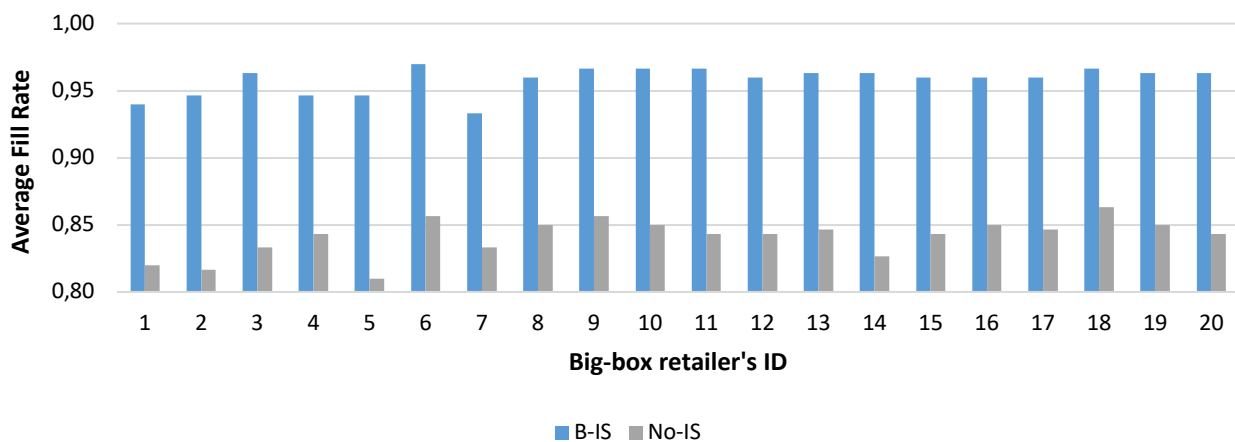

*Figure 13. Big-box retailers' average fill rate in the two scenarios*

The results of the non-parametric Mann-Whitney test on the fill rate of the 20 retailers (see Table 10) shows that a significant difference is achieved. If we consider as null hypothesis the following $H_0$ "the medians of the retailers' fill rates are equal" and as alternative hypothesis that "the median at B-IS is greater than the median at No-IS", we obtain a p-value (adjusted for ties) equal to <0,005, therefore we can reject $H_0$ and conclude that the fill rates are different from each other.

*Table 10. Non-parametric Mann-Whitney test for the retailers' fill rate*

| Scenario | N | Median | Difference | Lower Bound for Difference | W-Value | P-Value |
|---|---|---|---|---|---|---|
| B-IS | 20 | 96,17% | 11,67 | 11,33 | 610,00 | 0,000 |
| No-IS | 20 | 84,33% | | | | |

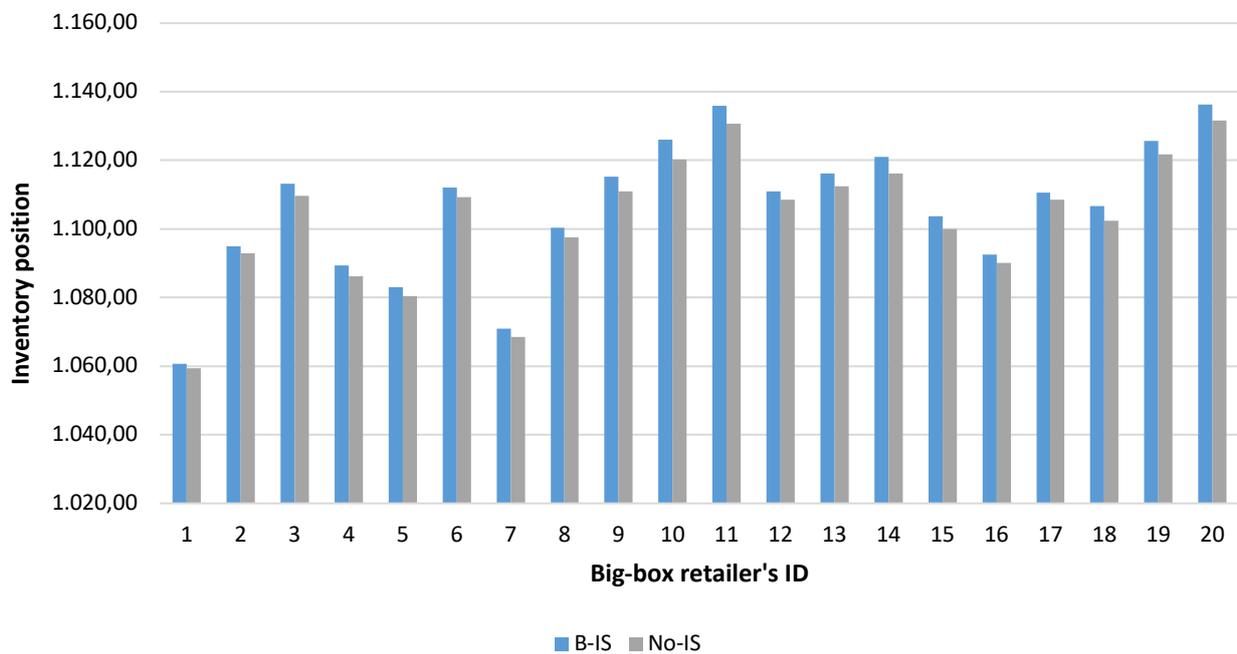

*Figure 14. Big-box retailers' inventory position in the two scenarios*

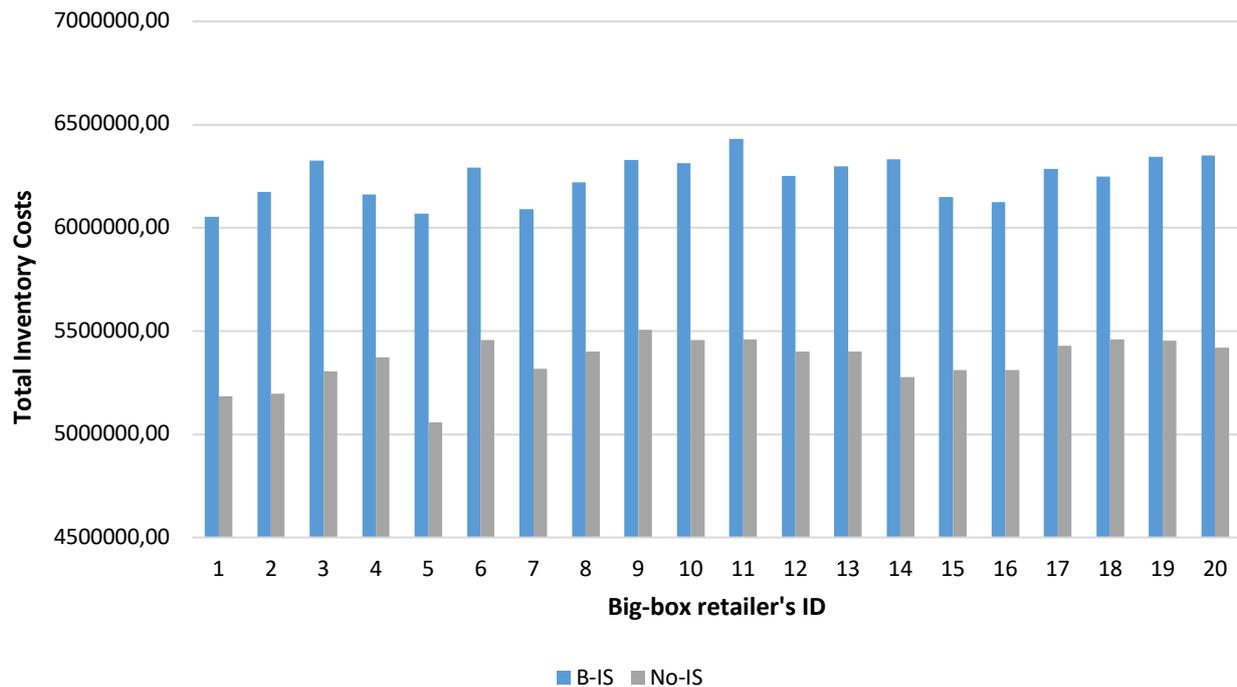

*Figure 15. Big-box retailers' total inventory costs in the two scenarios*

## 4. Discussion

Some insights about the results of the application study are provided to answer the two research questions that have driven the present work.

### 4.1. Is blockchain a convenient instrument for companies operating in a supply chain?

As expected, supply chain companies perform better in the B-IS scenario than in the No-IS one. Evident is the significant increase (as confirmed by the Mann-Whitney test) of the average fill rate, i.e. the number of orders satisfied versus the number of total orders, for both the wholesalers (Figure 8) and the big-box retailers (Figure 13). This can be explained by the fact that the in the No-IS scenario, companies do not trust each other and they can rely only on the history of purchase orders that a specific wholesaler has received by the retailers to make their forecasts. In the B-IS scenario, forecasts made by the wholesalers are more precise in the B-IS scenario as they leverage on actual data (lead time demand data are updated every day on the blockchain) shared by the big-box retailers. This has been made possible because trust is guaranteed by the blockchain, which acts as a certification agent and provides the state of truth and trust for the information exchanged. Each company can have a wallet in it and use the software connector module to certify the origin, the integrity and invariability of data it has just received by an untrustworthy partner through the hash sum stored in the blockchain.

Thanks to a better service level offered by the wholesalers (i.e. an increased fill rate) enabled by the blockchain-based information sharing and higher trust in shared data, the retailers are able to fulfill the market demand in a significantly improved manner. As a consequence of a better fill rate for the wholesalers, the inventory position – and therefore the costs – increase (see Figure 9 and Figure 10) but revenues increase as well (Figure 6) due to an improved forecasting and planning. Indeed, the wholesalers order more from the manufacturers, have a higher inventory position but they also sell more in the B-IS scenario compared to the No-IS scenario. Since the increase of the wholesalers' revenues compensate for the rise of their inventory costs, their profit margins remain nearly stable (see Figure 7). We can conclude that the use of blockchain to certify the origin, the integrity and invariability of data shared by the retailers significantly impact on the wholesalers' operational performance, but it does not significantly affect the economic benefits that they could achieve.

On the other side, although the big-box retailers address the market demand better than before (the fill rate is higher in the B-IS scenario, see Figure 13), their inventory position remains almost the same in both scenarios (only a 0,41% increase is observed in B-IS). The total inventory costs undergo only a 17,18% average increase that does not compromise the profit margins of the retailers that, in turn, are subject to a sharp increase thanks to the use of the blockchain. Hence, results show a reverse effect on the big-box retailers' performance: while their operational performance is not changing significantly in terms of inventory position, the higher fill rate generates significantly higher profit margins. Economic benefits can be therefore achieved by the big-box retailers that share supply chain related relevant data with their stakeholders and use blockchain to ensure data origin, the integrity and invariability.

It can be concluded that in a supply chain where companies do not trust each other or confidential data cannot be shared with the whole members of the supply chain, blockchain is a valuable technology that provides the conditions for a better operational performance (e.g. increased number of satisfied orders and higher service rate) of the whole supply chain. The significant economic benefits achieved by the retailers represent a notable driver to persuade them to adopt blockchain technology. On the other hand, suppliers would be encouraged by the mitigation of the bullwhip effect that allow them to better plan the processes and to achieve a higher operational performance. Results prove that blockchain technology is a cost-convenient instrument to overcome collaboration and trust issues in a supply chain and to minimize the negative consequences of information asymmetry over the echelons of a supply chain.

## 4.2. Are companies discouraged from sharing inaccurate or counterfeit data?

Despite the application study shows quantitatively which are the benefits that companies could achieve in a blockchain-enabled supply chain environment before its implementation in a real context, data that companies send to the blockchain can be still counterfeited or be inaccurate in advance, which may damage the whole supply chain operational performance. This situation would also have another ripple effect: if inaccurate or counterfeit data are shared, companies will no longer consider blockchain as a secure point of truth and trust and be prone to implement it. In this sense, inappropriate behavior should be discouraged. In the proposed application study, the only agents (supply chain members) who share information (lead time demand) on the blockchain are the big-box retailers. This information is made available one level upstream to the wholesalers with certain levels of visibility (retailers may decide to let only those wholesalers that ask for supply to access their information about lead time demand). As confirmed by the Mann-Whitney tests, only the retailers have obtained significantly higher margin profits, while the wholesalers' margin profits are similar in the two scenarios. We can conclude that only those supply chain members who share information obtain relevant benefits. Most certainly, when all the entities in the same level (all the retailers or all the wholesalers) share their information about the market demand, the greater advantages are obtained. Similarly, if the wholesalers strive to obtain similar advantages, this application shows that they need to share information with the level upstream (i.e. manufacturers in this case). Although people may argue that companies using the blockchain may use it in a malicious way by entering incorrect data about the lead time demand, this application provides material for interesting insights. Indeed, it is not convenient for the retailers to share wrong or incorrect data with the partners of the supply chain because it is their own performance that would be affected negatively by this inappropriate behavior. Therefore, the opportunity to increase their performance together with the features of immutability and traceability of the information in the blockchain network would be a deterrent for any misconduct by one of the supply chain members. Only when all the entities in the supply chain share correct information by leveraging on the blockchain, the supply chain itself can achieve the maximum competitiveness and performance.

## 4.3. Considerations about the blockchain costs

As far as the blockchain costs and maintainability is concerned, the investment is worthy as the system allows to register quasi real-time the hash sum in the distributed ledger (the transaction storing time is very low) and such information is immutable, thus providing the members upstream (the wholesalers) with a trustworthy way to check the data authenticity in a low-trust environment. The costs of maintaining such a network is negligible (the 0,003%) compared to the total costs (total inventory costs and blockchain costs) of the companies. Since the transaction storing time may increase when lots of data (coming from example from Internet of Things devices spread over the entire supply chain) will be exchanged and stored in the blockchain, the cost of using the blockchain may be no longer negligible. A way to overcome this issue is to share on the blockchain only aggregate data and preprocessed information (instead of raw data) that are crucial for the supply chain performance. For example, the big-box retailers should not share real-time on the blockchain the time at which every customer arrive at one of the stores, but at the end of the day, information about the lead time demand can be shared (as in the context of this application study). In the perspective of a more intense use of the blockchain, this expedient allows to keep the transaction storing time and the cost of using the Ethereum blockchain as low as possible.

### 4.4. Considerations about the developed solution

From a technological point of view, one of the main barriers to the use of blockchain in several contexts is the lack of plug&work solutions on the market that enable the enterprises' information systems to easily connect to the blockchain. The development of the software connector module here proposed is a first step towards a set of tools or low-cost solutions that provide companies the possibility to deal with data, export them and send them to other modules, applications or systems through interfaces that use (among others) REST web services.

The system has been proved to be very powerful to assess the potential benefits of blockchain technology in low-trust environments, such as a supply chain. The role of blockchain as a certification agent resulted to be successful in the perspective of the supply chain operational and economic performance assessed by means of the simulation model and fits the need for companies to have a decentralized entity that ensure the authenticity, validity and integrity of data stored off-chain, especially when data are confidential, or when data access rules or privacy must be respected or when entities in the supply chain do not trust each other (e.g. they do not know enough each other or they see themselves as competitors). The integration of a Java-based simulation model that faithfully recreates a real-world system (e.g. supply chain, production system, smart city, etc.) with an Ethereum-like blockchain through the ad-hoc developed software connector module may provide the prerequisite to investigate further the possible advantages of the distributed ledger technology, its potential role and implications for the real system.

### 5. Conclusions

Organizations in the same supply chain perceive data and information accuracy as a crucial factor of their performance but they are often averse in providing or using information when they do not trust each other. Despite companies may have access to the supply chain partners' data, trust issues still exist because they might mislead deliberately or unconsciously the supply chain partners with inaccurate, wrong or counterfeit information that does not reflect the real data. Companies are looking for methods and plug&work tools that enable them to share information in a secure way and check the origin, authenticity and integrity of data over time so that they can make more reliable and trustworthy plans and forecasts. Blockchain technology is a perfect solution to these problems since it establishes a single, immutable record of data that can be viewed by anyone has the right to access it and that cannot be altered. While the rise of blockchain in finance has been extremely rapid, enterprises claim to have little knowledge about blockchain due to its novelty and to the lack of use cases and application studies in literature that describe blockchain's potential benefits.

The present paper proposed the role of blockchain as a certification entity that guarantees the origin, authenticity and integrity of data stored off-chain. To this end, a software connector module has been developed to enable the interaction between an enterprise information system to an Ethereum-like blockchain. In order to show quantitatively the benefits that companies can achieve in a blockchain-enabled supply chain, a supply chain simulation model has been developed to carry out "what-if" scenario analysis and integrated with the blockchain via the same software connector through REST web services to serve as a replica of a real supply chain. The application study shows that blockchain technology is a convenient instrument to overcome collaboration and trust issues in a supply chain, to increase the supply chain overall performance, to minimize the negative consequences of information asymmetry over the echelons of a supply chain but also to discourage companies from any misconduct (e.g. counterfeiting data or low data accuracy).

Since supply chain management research on blockchain is still in its infancy, it is worth to start looking into possible applications and benefits that may convince supply chain managers to adopt this technology and operate in an environment where everyone trusts each other. This work fosters the exploration of further supply chain phenomena by using a blockchain-enabled simulated supply chain, which represents the perfect test environment to explore the real advantages of blockchain technology.